\documentstyle[seceq,epsf,twoside]{ptptex}
\notypesetlogo  
\title{
\hfill{\normalsize\vbox{\hbox{DPNU-00-28}  }}\\
\ \\
Pi Pi Scattering and Scalar Mesons 
in an Effective Chiral Lagrangian\footnote{%
Talk given at YITP workshop ``Possible existence of 
the $\sigma$-meson and its implications to hadron physics''
(June 12--14, 2000, Kyoto).
This talk is based on the work done in collaboration with
Prof.~J.~Schechter and Dr.~F.~Sannino in 
Refs.~\ref{ref: Harada-Sannino-Schechter:1}
and \ref{ref: Harada-Sannino-Schechter:2}.%
\nocite{Harada-Sannino-Schechter:1,Harada-Sannino-Schechter:2}
}}
\author{%
Masayasu {\sc Harada}\footnote{%
e-mail: {\tt harada@eken.phys.nagoya-u.ac.jp}%
}}
\inst{%
 Department of Physics, Nagoya University\\
Nagoya 464-8602, Japan
}
\abst{%
In this talk I summarize recently proposed mechanisms to understand
$\pi\pi$ scattering to 1\,GeV in an effective chiral Lagrangian.
The Lagrangian includes higher resonances in addition to pions
consistently with the chiral symmetry.
Iso-spin zero $S$-wave partial wave amplitude is reproduced up till
about 1.2\,GeV by including a pion self-interaction and resonant pole
exchanges of $\rho$, $f_0(980)$ and $\sigma$ derived from the
effective chiral Lagrangian.
The best fit shows that $\sigma$ has a mass of around 560\,MeV and a
width of about 370\,MeV.
}

\begin{document}
\maketitle

\setcounter{tocdepth}{4}

\section{Introduction}

QCD is known to be the fundamental theory of the strong
interaction. 
However, it is very difficult to reproduce experimental data
directly from QCD.
One clue to study low energy properties of QCD
is given by the structure of the chiral symmetry,
which approximately exists in the QCD Lagrangian and
is broken by the strong interaction of QCD.
Another clue is given by the $1/N_{\rm C}$ expansion to 
QCD.\cite{1/Nc}

The $\pi\pi$ scattering has been studied as an important test of the
low energy properties of QCD.
The experimental data in 
the low energy region near $\pi\pi$ threshold can be
reproduced by using the information from
chiral symmetry.
This situation is easily understood by using a chiral Lagrangian
which includes pions only.
In addition, by including higher derivative terms together with
one-loop effects, the applicable energy region is enlarged.
This systematic low energy expansion is called the chiral perturbation
theory.\cite{ChPT}

In the higher energy region, however,
the one-loop amplitude of the chiral perturbation theory
violates the unitarity around 
$400-500$\,MeV in the $I=0$ $S$-channel.\cite{Gasser-Meissner}
For the $P$-wave amplitude,
we have the $\rho$ meson, and the chiral perturbation theory
may break down at the resonance position.
The explicit inclusion of resonances in the high energy region
easily reproduces the amplitude.
In such a case, however, it is important to consider clues
to QCD other than the chiral symmetry.

In the large $N_c$ limit QCD becomes a theory of weakly interacting
mesons, and the $\pi\pi$ scattering is expressed by infinite sum of
tree graphs.
However, we cannot actually include infinite number of resonances.
Moreover, the forms of interactions are not fully determined in
the large $N_{\rm C}$ limit.
Nevertheless, some encouraging features were previously found in an
approach which truncated the particles appearing in the effective
Lagrangian  to those with masses up to an energy slightly greater than
the range of interest.\cite{Sannino-Schechter}
Then we constructed a resonance model, where
we truncated the theory, and included
particles with masses up to an energy slightly greater than the range
of interest.
Moreover, the chiral symmetry was used to restrict forms of
interactions,
i.e., the effective Lagrangian was constructed by
using the information of chiral symmetry.
This seems reasonable phenomenologically and is what one
usually does in setting up an effective Lagrangian. 

In this talk I concentrate on the energy region below 1\,GeV.
For the established resonances lighter than 1\,GeV,
$\rho$ and $f_0(980)$ are contained in the particle data group (PDG)
list\cite{ParticleDataGroup:00} (see Table~\ref{tab: particles}).
\begin{table}
\begin{center}
\begin{tabular}{|c||c||c|c|} \hline
 &$I^G(J^{PC})$ & $M$(MeV) & $\Gamma$(MeV) \\ 
\hline \hline
$\sigma(400-1200)$ & $0^+(0^{++})$  & 400$-$1200 & 600$-$1000 \\
$\rho(770)$   & $1^+(1^{--})$  & 769.3   & 150.2  \\
$f_0(980)$   & $0^+(0^{++})$  & 980     & 40$-$100 \\
\hline
\end{tabular}
\end{center}
\caption[]{
Resonances included in the $\pi\pi\rightarrow\pi\pi$ channel as
listed in the PDG\cite{ParticleDataGroup:00}. 
}
\label{tab: particles}
\end{table}
However, the width of $f_0(980)$ is not well determined.
Moreover, the existence of a light scalar $\sigma$ is suggested by
several authors.\cite{sigma}
Here I will determine these resonance parameters by fitting to the
$I=0$ $S$-wave $\pi\pi$ scattering amplitude.

This talk is organized as follows.
In section~\ref{sec: model} I will show the interactions among the
higher resonances and two pions, which are derived from an effective
chiral Lagrangian.
Section~\ref{sec: fit} is the main part of this talk, where I will
show how to regularize the amplitude, and fit the resonance
parameters to the experimental data of the $I=0$, $J=0$ partial wave
amplitude.
Finally, a summary is given in section~\ref{sec: summary}.

\section{Resonance Model}
\label{sec: model}

In this section I will show the interactions of the higher resonances, 
listed in Table~\ref{tab: particles}, with two pions.
These interactions are derived from an effective chiral Lagrangian 
which includes the higher resonances
consistently with the chiral symmetry.
The starting effective chiral Lagrangian includes pions through the
non-linear realization of the chiral symmetry breaking.

First I include the vector meson as a gauge field of chiral
symmetry\cite{Kaymakcalan-Schechter},
which is equivalent to the hidden local gauge 
method
(See, for a review, Ref.~\ref{ref: Bando-Kugo-Yamawaki:PRep}.)
\nocite{Bando-Kugo-Yamawaki:PRep}
at tree level.
This leads to the following $\rho\pi\pi$ interaction:
\begin{equation}
{\cal L}_\rho = g_{\rho\pi\pi} \vec{\rho}_\mu \cdot
\left( \partial^\mu \vec{\pi} \times \vec{\pi} \right)
\ ,
\end{equation}
where $g_{\rho\pi\pi}$ is the $\rho\pi\pi$ coupling constant.

Next, I include scalar resonances, $\sigma$ and $f_0(980)$.
These are iso-singlet fields.  Inclusion of an iso-singlet scalar
field consistently with the chiral symmetry leads to the following
interaction among one scalar and two pions:
\begin{equation}
{\cal L}_{f}=-\frac{\gamma_f}{\sqrt{2}}\; f\;
\partial_{\mu}\vec{\pi}\cdot\partial_{\mu}\vec{\pi}
\qquad \left( f=\sigma\,,f_0(980) \right)
\ .
\label{la:sigma}
\end{equation}
Here I should note that the
chiral symmetry requires derivative-type interactions
between the scalar and pseudoscalar mesons.

\section{Fit to $\pi\pi$ Scattering to 1 GeV}
\label{sec: fit}

In this section, I will calculate the $I=0$
$S$-wave $\pi\pi$ scattering
amplitude by including resonances as explained in the previous
section.

The most problematic feature involved in comparing the leading
$1/N_{\rm C}$ amplitude
with experiment is that it does not satisfy unitarity.
Since the mesons have zero width in the large $N_{\rm C}$ limit,
the amplitude diverges at the resonance position.
Thus in order to compare the $1/N_{\rm C}$ amplitude with experiment
we need to regularize the resonance contribution.
Here let me summarize the regularizations.

Ordinary narrow resonances such as $\rho$ are regularized by
including the width in the denominator of the propagator 
(the Breit-Wigner form):
\begin{equation}
\frac{M\Gamma}{M^2-s-iM\Gamma}\ .
\label{Breit-Wigner}
\end{equation}
This is only valid for a narrow resonance in a region where the
{\it background} is negligible.
Note that the width in the denominator is related to the coupling
constant.

For a very broad resonance
there is no guarantee that such a form is correct. 
A suitable form turned out to be of the type:
\begin{equation}
\frac{M G}{M^2-s-iMG^\prime}\ ,
\label{sigma-propagator}
\end{equation}
where the parameter $G^\prime$ is a free
parameter. This $G^\prime$ is not related to the coupling constant.

Even if the resonance is narrow, the effect
of the background may be rather important.
This seems to be true for the case of $f_0(980)$.
Demanding local unitarity in this case yields a partial
wave amplitude of the well known form:\cite{Taylor}
\begin{equation}
\frac{e^{2i\delta}M\Gamma}{M^2-s-iM\Gamma}+e^{i\delta}\sin \delta\ ,
\label{rescattering}
\end{equation}
where $\delta$ is a background phase (assumed to be slowly varying).
I will adopt a point of view in which this form is regarded as a kind
of regularization of the model. Of course, non zero $\delta$
represents a rescattering effect which is of higher order in
$1/N_{\rm C}$.
The quantity $\displaystyle{e^{2i\delta}}$, taking $\delta=constant$,
can be incorporated into the squared coupling constant connecting the
resonance to two pions.
In this way, crossing symmetry can be preserved.
The non-pole background term in Eq.~(\ref{rescattering}) and hence
$\delta$ is to be predicted by the other pieces in the effective
chiral Lagrangian. 

Another point which must be addressed in comparing the leading
$1/N_{\rm C}$ amplitude
with experiment is that it is purely real away from the singularities.
The regularizations mentioned above do introduce some imaginary pieces
but these are clearly more model dependent.
Thus it seems reasonable to compare the real part of the predicted
amplitude with the real part of the experimental amplitude.

Let me start from the {\it current algebra} + $\rho$ contribution.
The predicted curve is shown in Fig.~1 of 
Ref.~\ref{ref: Harada-Sannino-Schechter:1}.
Although the introduction of $\rho$ dramatically
improves unitarity up to about $2$\,GeV, $R^0_0$ violates unitarity to
a lesser extent starting around $500$\,MeV.
To recover unitarity, we need a negative contribution to the real part
above this point, while below this point a positive contribution is
preferred by experiment.
Such behavior matches with the real part of a typical
resonance contribution.
The resonance contribution is positive in the energy region below its
mass, while it is negative in the energy region above its mass.
Then I include a low mass broad scalar resonance, $\sigma$.
The $\sigma$ contribution to the real part of the amplitude component
$A(s,t,u)$ is given by
\begin{equation}
A_{\sigma}(s,t,u)=
\frac{\gamma_\sigma^2}{2}
\frac{(s-2m_\pi^2)^2}{M_\sigma^2-s - i M_\sigma{G^\prime}}\ ,
\label{eq:sigma}
\end{equation}
where the factor $(s-2m_{\pi}^2)^2$ is due to the derivative-type
coupling required for chiral symmetry in Eq.~(\ref{la:sigma}).
$G^{\prime}$ is a parameter which we introduce to regularize the
propagator.
It can be called a width, but it turns out to be rather large so that,
after the $\rho$ and $\pi$ contributions are taken into account, the
partial wave amplitude $R^0_0$ does not clearly display the
characteristic resonant behavior.

A best overall fit is obtained with the parameter
choices; $M_{\sigma}=559$\,MeV, 
$\gamma_{\sigma}=7.8$\,GeV$^{-1}$ and $G^{\prime}=370$\,MeV.
The result for the
real part $R^0_0$ due to the inclusion of the 
$\sigma$ contribution along
with the $\pi$ and $\rho$ contributions is shown in Fig.~1.
It is seen that the unitarity is satisfied and there is a
reasonable agreement with the experimental 
points\cite{Alekseeva,Grayer} up to about $800$\,MeV.

\begin{figure}[htbp]
\noindent
\begin{minipage}[t]{0.46\textwidth}
\begin{center}
\epsfxsize=0.92\textwidth
\ \epsfbox{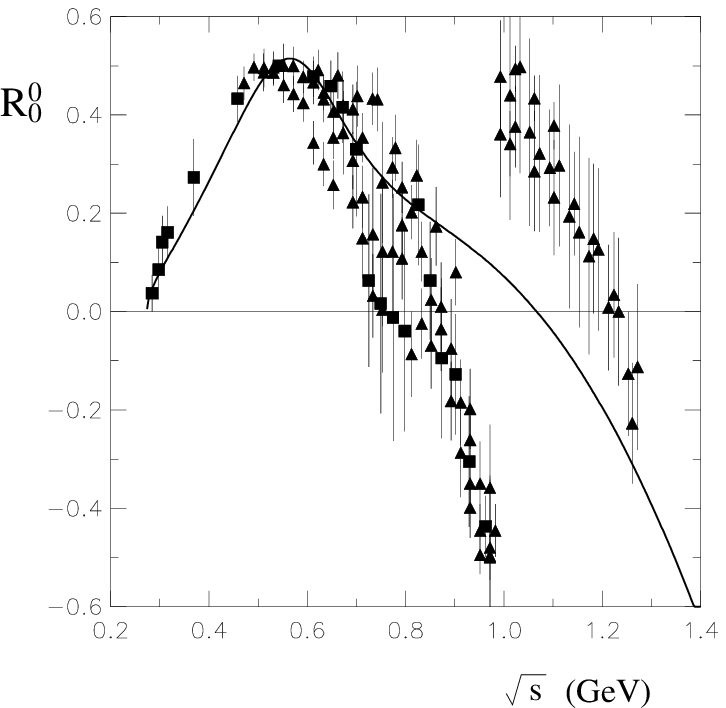}
\end{center}
\addtocounter{figure}{1}
\noindent
\footnotesize
Figure~\thefigure:
\label{figura2}
The solid line is the {\it current algebra}
$+~\rho+\sigma$ result for $R^0_0$.
The experimental points, in this
and succeeding figure, are extracted from the phase shifts
($\Box$: Ref.~\ref{ref: Alekseeva}\nocite{Alekseeva},
$\triangle$: Ref.~\ref{ref: Grayer}.\nocite{Grayer}).
\end{minipage}
\hspace{\columnsep}
\begin{minipage}[t]{0.46\textwidth}
\begin{center}
\epsfxsize=0.92\textwidth
\ \epsfbox{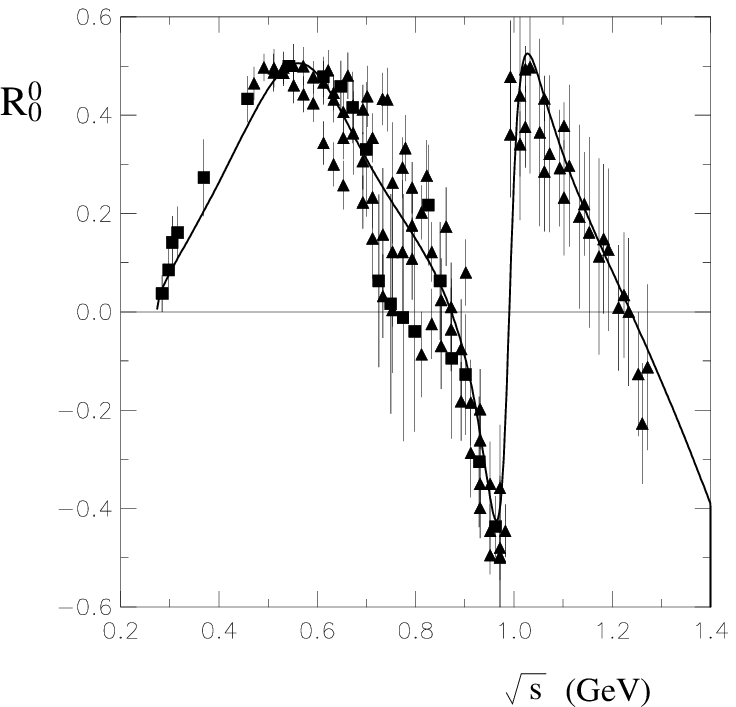}
\end{center}
\addtocounter{figure}{1}
\noindent
\footnotesize
Figure~\thefigure:
The solid line is the {\it current algebra}
$+~\rho~+~\sigma~+~f_0(980)$ result for $R^0_0$ obtained by assuming
the values in Table~\ref{tab: result}
for the $\sigma$ and $f_0(980)$ parameters.
\label{figura4}
\end{minipage}
\end{figure}

Next, let me consider the 1\,GeV region.
Reference to Fig.~1 shows that the experimental data for
$R^0_0$ lie considerably lower than the $\pi+\rho+\sigma$ contribution
between $0.9$ and $1.0$\,GeV and then quickly reverse sign above this
point.
This is caused by the existence of $f_0(980)$.
As we can see easily, a naive inclusion of $f_0(980)$ does not
reproduce the experimental data,
since the real part of the typical resonance form gives a positive
contribution in the energy region below its mass, while it gives a
negative contribution in the energy region above its mass.
However, we need a negative contribution below 1\,GeV
and a positive contribution above 1\,GeV.

As I discussed around Eq.~(\ref{rescattering}), the effect of the
background is important in this $f_0(980)$ region.
In this case the background is 
given by the $\pi+\rho+\sigma$ contribution.
Figure~1 shows that the real part of the background is
very small so that the background phase $\delta$ 
in Eq.~(\ref{rescattering}) is expected to be roughly 90$^\circ$.
This background effect generates an extra minus sign in front of the
$f_0(980)$ contribution, as we can see from Eq.~(\ref{rescattering}).
Thus $f_0(980)$ gives a negative contribution below the resonance
position and gives a positive contribution above it.
This is exactly what is needed to bring experiment
and theory into agreement up till about $1.2$\,GeV.

A best fit of our parameters to the experimental data results in the
curve shown in Fig.~2.
Only three parameters $\gamma_{\sigma}$, $G'$ 
and $M_\sigma$ are essentially free.
The others are restricted by experiment.
Since the total width of $f_0(980)$ has a large uncertainty 
(40 -- 100\,MeV in PDG list),
we also fit this.
In addition we have considered the precise value of $M_{f_0}$ to be a
parameter for fitting purpose.
The best fitted values are shown in Table~\ref{tab: result} together
with the predicted background phase $\delta$ and the $\chi^2$ value.
The predicted background phase is seen to be close to 90$^\circ$,
and the low lying sigma has a mass of around 560\,MeV and a width of
about 370\,MeV.
\begin{table}[hbtp]
\begin{center}
\begin{tabular}{|c|c|c|c|c|c|c|}
\hline
$M_{f_0(980)}$ & $\Gamma_{f_0(980)}$ & $M_\sigma$ & $G'$ 
 & $\gamma_{\sigma}$ & $\delta$ (deg.) & $\chi^2$ \\
\hline
987 & 64.6 & 559 & 370 & 7.8 & 85.2 & 2.0 \\
\hline
\end{tabular}
\end{center}
\caption[]{The best fitted values of the parameter together with the
predicted background phase $\delta$ and the $\chi^2$ value.
The units of $M_{f_0(980)}$, $\Gamma_{f_0(980)}$, $M_\sigma$ and $G'$
are MeV and that of $\gamma_{\sigma}$ is GeV$^{-1}$.
}
\label{tab: result}
\end{table}

\section{Summary}
\label{sec: summary}

In this talk I showed main mechanisms of the analysis done in
Ref.~\ref{ref: Harada-Sannino-Schechter:1}:
(1) motivated by the large $N_{\rm C}$ approximation to QCD, we
include resonances with masses up to an energy slightly greater
than the range of interest, and use the chiral symmetry to restrict
the forms of the interactions;
(2) the {\it current algebra} + $\rho$ contribution 
violates the unitarity around 560\,MeV region but it is recovered by
including the low mass broad resonance
$\sigma$\cite{Sannino-Schechter};
(3) the $\pi$ + $\rho$ + $\sigma$ contribution gives an important
background effect to the $f_0(980)$ contribution, i.e., the sign in
front of the $f_0(980)$ contribution is reversed by the background
effect.
The third mechanism, which leads to a sharp dip in the $I=J=0$ partial
wave contribution to the $\pi\pi$-scattering cross section, can be
identified with the very old {\it Ramsauer-Townsend} 
effect~\cite{Schiff} which concerned the scattering of 0.7\,eV
electrons on rare gas atoms.
The dip occurs because the background phase of $\pi/2$ causes the
phase shift to go through $\pi$ (rather than $\pi/2$) at 
the resonance position.
(Of course, the cross section is proportional to 
$\sum_{I,J}^{~}(2J+1) \sin^2(\delta^J_I)$.)
This simple mechanism seems
to be all that is required to understand the main feature of $\pi\pi$
scattering in the $1$\,GeV region.

\acknowledgements

I would like to thank Prof.~Joe~Schechter and Dr.~Francesco~Sannino
for collaboration in Refs.~\ref{ref: Harada-Sannino-Schechter:1}
and \ref{ref: Harada-Sannino-Schechter:2}.
I would like to thank Prof.~Joe~Schechter for his visit to Japan and 
stimulate discussions.
I also would like to thank
the organizers of this workshop for giving me an
opportunity to present this talk.
This work is supported in part by Grant-in-Aid for Scientific Research
(A)\#12740144.

\end{document}